\documentclass[12pt]{article}
\usepackage{epsfig,amsfonts}

\begin{document}

\title{A model for nuclear matter fragmentation: phase diagram and cluster
distributions}

\author{J.M. Carmona$^a$, J. Richert$^b$, A. Taranc\'on$^a$\\[0.5em] 
$^a$ {\small Departamento de F\'{\i}sica Te\'orica,}
{\small Universidad de Zaragoza,}\\
{\small Pedro Cerbuna 12, 50009 Zaragoza (Spain)}\\[0.3em]
$^b$ {\small Laboratoire de Physique Th\'eorique,}
{\small Universit\'e Louis Pasteur,}\\
{\small 3, rue de l'Universit\'e, 67084 Strasbourg Cedex (France)}\\[0.3em]
{\small \tt carmona@sol.unizar.es}\\
{\small \tt richert@lpt1.u-strasbg.fr}\\
{\small \tt tarancon@sol.unizar.es}\\[0.5em]}

\date{\today}

\maketitle

\begin{abstract}
We develop a model in the framework of nuclear fragmentation at 
thermodynamic equilibrium which can be mapped onto an Ising model with
constant magnetization. We work out the thermodynamic properties of the
model as well as the properties of the fragment size distributions.
We show that two types of phase transitions can be found for high density
systems. They merge into a unique transition at low density. An analysis
of the critical exponents which characterize observables for different
densities in the thermodynamic limit shows that these transitions look like
continuous second order transitions.
\end{abstract}

\vskip 5 mm

\vfill

\medskip

Phase transitions.
Monte Carlo.
Ising model.
Kawasaki dynamics.

\thispagestyle{empty}

\newpage

\section{Introduction}

The fragmentation of excited nuclei is a complex, time-dependent process.
There exists however experimental evidence if no proof that the decaying
system can be treated as a thermodynamically equilibrated assembly of
interacting particles, at least in specific experimental circumstances
\cite{Bord}. If this is the case, the fragmentation process evolves in time 
by following trajectories in the $(\rho,T)$ phase diagram ($\rho$ is
the density and $T$ the temperature). It starts at some initial point and
ends at a final point when the freeze-out stage is reached. During the time
interval in which this goes on, the initial excited system breaks into 
fragments and (or) evaporates particles.

As far as such a scenario corresponds to a physical realization of the
process, one is left with two types of experimentally relevant physical 
observables. The first set concerns the thermodynamic properties of the
expanding system (energy, temperature, pressure and correlations between 
these quantities, possible existence of phase transitions). The second set
concerns final fragment size distributions and related observables.

The latter observables have been rather extensively studied in numerous 
experiments, showing remarkable properties \cite{Schu} which were analyzed 
by means of different types of models. These analyses lead
to the conclusion that the system may undergo a second order transition
which lies in the universality class of bond percolation in three
dimensions \cite{Zhen}.

The thermodynamic properties of fragmenting nuclei are less well known
from the experimental point of view. There are indications for the existence
of two phases which could be analogous to liquid and gas in macroscopic
systems. The order of the transition between the two phases is not clearly
established either~\cite{Poch,Haug,Ma}. Successful multifragmentation
models~\cite{gross1,Gross,Bondorf} which are able to reproduce a large amount
of experimental data predict a first order transition.

Running parallel to the experimental efforts in order to characterize the
behaviour of the system, realistic theoretical models should aim and be able
to describe both aspects, the thermodynamic properties as well as the
characteristics of the fragment size distributions in a consistent way.
This has already been attempted in recent work in the framework of the lattice
gas model (LGM) \cite{campi,pan1,pan2,pan3} 
and to some extents by means of homogeneous cellular models \cite{Rich}.

It is the aim of the present paper to present a detailed analysis of both
properties by means of a simple though quite realistic model, to identify
the phases in which the system can exist and the order of the 
corresponding phase transitions.

The paper is arranged in the following way. In section 2 we describe the
model. We emphasize its kinship with other models and its
specific features. In section 3 we first define the observables which we
consider in the sequel and work out the phase diagram which links the density
to the temperature. We then analyze the properties of the phase transitions
which appear in the system in section 4. We summarize, conclude, present
and discuss further possible developments in section 5.

\section{Spin models and fragmentation}

\subsection{The lattice gas model and the Ising model}

The lattice gas model (LGM) describes the motion of $N=\rho\cdot V$ 
particles inside 
a three-dimensional cubic lattice, where $\rho$ is the average density,
$0\leq\rho\leq 1$, and $V$ the number of sites. Particles which occupy sites
in this volume interact only when they 
are nearest-neighbours. If $n_i=0,1$ are the site occupation
numbers (empty/occupied), and $\mu$ is the chemical potential, 
the Hamiltonian reads
\begin{equation}
H_V=-4J\sum_{\langle i,j\rangle\subset V}n_i n_j+\mu\sum_{j\in V}n_j,
\label{ham1}
\end{equation}
where $\langle i j\rangle$ are nearest neighbour cells.

The formulation presented in eq. (\ref{ham1}) corresponds to the
grand canonical ensemble. The microcanonical ensemble is the most natural
ensemble to be used when describing nuclear fragmentation since nuclei are
closed systems with fixed energy and number of particles. 
It is usually assumed that the thermodynamic results are the same
independently of the formalism used. Of course this is not true with finite
systems, because microcanonical, canonical and grand canonical ensembles
differ on the level of fluctuations. But it has recently been 
argued~\cite{gross2}
that even in the thermodynamic limit, microcanonical and (grand)canonical
ensembles are not equivalent (for example, at a first order transition).
We shall explicitely show in this paper that in fact the canonical and
grand canonical formulations of the LGM are not equivalent. This
happens because there are points in the phase diagram which are accessible
when using one formalism, but are not defined (they are not points of
thermodynamic equilibrium) when using another one. As we will see, this causes
the order of the transition to be different in the two formalisms. This 
fact leads us naturally to ask about the behaviour of the system in the
microcanonical ensemble. This is an open problem, but as a first step
we will consider here only the detailed study of the canonical
formulation to compare it with the well-known grand canonical version of
this model. In fact, for sytems with finite volume, the microcanonical
ensemble turns out to be very efficient in the study of 
phase transitions~\cite{gross3,lukk}.

This simple model can be numerically studied, 
its grand partition function being proportional to the canonical partition 
function of the Ising model with an external magnetic field \cite{LY}.
In fact, setting
\begin{equation}
\sigma_j=2 n_j-1,
\end{equation}
the Hamiltonian (\ref{ham1}) becomes
\begin{eqnarray}
H_V &=& -J\sum_{\langle i,j\rangle\subset V}\sigma_i\sigma_j+
\left(\frac{\mu}{2}-6J\right)\sum_{j\in V}\sigma_j+
\mathrm{const}(V) \\ \nonumber
&\equiv& -J\sum_{\langle i,j\rangle\subset V}\sigma_i\sigma_j-
h\sum_{j\in V}\sigma_j+\mathrm{const}(V),
\end{eqnarray}
where we reinterpret $\sigma_j=\pm 1$ as a spin variable,
and identify a spin up with an occupied site, and 
a spin down with an empty site. We introduce
\begin{equation}
M\equiv\frac{1}{V}\sum_j \sigma_j=
\frac{1}{V}(2N-V)=2\rho-1.
\label{density}
\end{equation}
$M$ is called the magnetization in the sequel. 
As it can be seen, the initial model defined 
through (\ref{ham1}) can indeed be mapped on an Ising model with a linear term 
corresponding to a fixed magnetic field $h$.

It is well known that the Ising model with magnetic field $h$ presents a
discontinuity in the magnetization $M(h,T)$ for $T<T_c$ at $h=0$, where 
$T_c$ is the critical temperature. That is, in the plane $(h,T)$ there exists
a line of first-order transitions $(h=0,\,0\leq T<T_c)$, and a second
order transition at $(h=0,T_c)$. Let us see how this is reflected in the
case of the LGM. Since $h$ plays essentially the role of 
the chemical potential in the grand canonical formulation of the LGM, there 
is a first order transition in the phase diagram $(\rho,T)$ of the LGM, 
which separates the homogeneous phase from the zone with the two ``liquid''
and ``gas'' phases. On this line, there exists a critical point at $\rho=0.5$,
$T=T_c$.

As described below, we aim to introduce a similar type of model 
in order to describe a finite, excited system of classical particles which
interact by means of a short range potential.
We characterize the internal structure of the system by means of
a proper definition of physical droplets.
Droplets are clusters of particles in which each particle has at least one
nearest neighbour. They exist because of two physical reasons: the attractive
nearest-neighbour interaction and the density of particles, 
which leads to usual percolation when this density is large enough.
However, such clusters are
not always stable, because they can break at zones of small connectivity
owing to the kinetic energy of the particles. The simple 
model (equivalently, the LGM) does not take the kinetic energy into account.
We can introduce this energy by means of another definition
which takes the energetics of the fragmentation into account~\cite{campi,pan1}.
Let us consider the stability of a cluster of $A$ particles. If the loss of
a particle implies the break of a single bond, then the excess of energy
will be
\begin{equation}
\frac{-p^2}{2m}+J,
\end{equation}
where $m$ is the mass of the particles of the LGM. The cluster is a stable
droplet if this excess is positive:
\begin{equation}
\frac{p^2}{2m}-J<0.
\end{equation}
This suggests to introduce a bond probability $p'_b\leq 1$ for the nearest
neighbour particles to belong to the same droplet:
\begin{equation}
p'_b=\mathrm{Prob}\left(\frac{p^2}{2m}-J<0\right),
\label{bondprev}
\end{equation}
where $p$ is distributed according to a Maxwell law, that is, the probability
for a particle to have a momentum $\mbox{\boldmath $p$}$ is
\begin{equation}
\mathrm{Prob}(\mbox{\boldmath $p$})=\frac{1}{[2\pi k_\mathrm{B}T]^{3/2}}
\,\exp\left(\frac{\mbox{\boldmath $p$}}{2mk_\mathrm{B}T}\right).
\end{equation}
One can see that the choice of the bond probability~(\ref{bondprev}) allows
to extend the argument to the case in which the loss of a particle of the
cluster means the break of several bonds~\cite{campi}. We obtain:
\begin{eqnarray}
p'_b&=&\mathrm{Prob}\left(p^2<2mJ\right)=1-\mathrm{Prob}\left(p^2>2mJ\right)
\nonumber \\
&=&1-\frac{4\pi}{[2\pi mk_\mathrm{B}T]^{3/2}}
\int_{\sqrt{2mJ}}^\infty e^{-p^2/2mk_\mathrm{B}T}p^2 \,\mathrm{d}p
\nonumber \\
&=&1-\frac{4}{\sqrt{\pi}}\int_{\sqrt{J/k_\mathrm{B}T}}^\infty
u^2 e^{-u^2} \, \mathrm{d}u.
\end{eqnarray}
Performing a Laguerre expansion, one finds~\cite{campi}
\begin{equation}
p'_b=1-0.911e^{-J/2k_\mathrm{B}T}-0.177\left(1-\frac{J}{k_\mathrm{B}T}\right)
e^{-J}{2k_\mathrm{B}T}+\cdots
\end{equation}
which shows the similarity between the bond probability 
definition~(\ref{bondprev}) and the Coniglio-Klein distribution of 
probability
\begin{equation}
p_b=1-e^{-2J/k_{\mathrm{B}}T}.
\label{bond}
\end{equation}
We will use this much simpler bond probability to simulate the kinetic
energy contribution. This definition of bond-correlated percolation has also
the advantage that the droplets percolate at the thermodynamic transition
point with the exponents of the Ising
model \cite{CoKl}. This is not true in general with the site-correlated
percolation given by the Ising clusters ($p_b=1$).

\subsection{Previous results}

The LGM has been introduced in order to describe fragmentation.
In refs. \cite{pan1,pan2} several analytical simplifications were made on 
the model, as well as numerical simulations taking the analogy with the
Ising model at fixed magnetic field. The authors used a definition of droplets 
which lead the thermal critical point to coincide with a percolation point.
Recently the authors of refs.~\cite{pan1,pan2} have also considered 
isospin-dependent interactions, and they have determined the
distribution of clusters together with other observables \cite{pan3}.
Similar studies can be found in ref.~\cite{chomaz}.

In ref. \cite{campi} it was noted that there exists a full line, not only
a single point, in the phase diagram of the LGM, for densities larger
than 0.5, where distributions of fragment sizes which obey a power law 
distribution are obtained.
On this line, which is very close to the Kert\'esz line \cite{Kert}, 
one finds a phase transition
because the droplets are defined with a physical prescription. 
Along this transition, there are no discontinuities for any
thermodynamic quantities (energy, specific heat,...). However, for some
nonlocal quantities, as clusters, a discontinuity appears.
The critical
exponents along this line (outside of the critical point) should give
some clue about the kinship with a percolation transition.

\subsection{Present model: Ising with fixed number\\of particles}

In all the numerical calculations mentioned above, the LGM is equivalent
to the Ising model with a constant magnetic field. This corresponds to
a fixed chemical potential. Hence the number of particles (and consequently,
the density) is conserved on the average in the framework of the 
grand canonical formalism.
For systems as small as nuclei, it is certainly appropriate to
consider the model with a fixed number of particles, which
corresponds to the canonical ensemble. 
As it was stated above, the nucleus is a closed system without heat bath.
Under these conditions, the microcanonical ensemble is the most appropriate
formalism to use, but we will leave the constraint of the energy
as an open problem and concentrate
on the crucial difference which exists between the canonical and the
grand canonical in this case. The reason is that the constraint of 
fixed number of particles can not
always be satisfied with the last formalism, in sharp contrast with the
situation in the former framework as we shall see.

The LGM in the canonical formalism is just the Ising 
model with fixed
magnetization (IMFM), here with fixed number of particles. We introduce
the following partition function
\begin{equation}
\mathcal{Z}_V=\sum_{[\sigma]}e^{\beta\sum_{\langle i j
    \rangle} \sigma_i\sigma_j}\,\cdot \delta(M-\hat M),
\label{model}
\end{equation}
where $\beta=J/k_{\mathrm{B}}T$ 
(we take $J/k_{\mathrm{B}}=1$ for simplicity), 
$\hat M\in [-1,1]$ is a fixed number, and 
the delta function constrains the configurations $[\sigma]$ which contribute
to the sum to those with fixed density, so that $M=\hat M$, where $M$
is defined in (\ref{density}) above.

As already stated above, the grand canonical formalism 
of the LGM is the Ising model with 
constant $h$. Of course an Ising model with a magnetic
field $h$ which produces a magnetization $\langle M\rangle$ at a certain
value of $\beta$ is equivalent in the thermodynamic
limit (that is, it
gives the same value for the different observables) to the model (\ref{model})
with magnetization $\hat M=\langle M\rangle$ and the same $\beta$. 
This is so because in the limit $V\to\infty$, the fluctuations in $M$
are negligible, and the distribution is a Dirac delta 
centered in $\langle M\rangle$.

But, reversely, the description of a model with fixed magnetization is not
necessarily equivalent to a model with fixed magnetic field in the grand
canonical ensemble:
in the LGM (we will refer from now on to the LGM as the model considered
in the grand canonical formalism), we obtain a certain mean magnetization
(or density, see eq. (\ref{density})) with a magnetic field $h$ at some
value of $\beta$. When we increase $\beta$, we have to lower $h$ 
in order to get
the same value of $\langle M\rangle$. 
It is always possible to find a value of $h$ which gives a fixed
$\langle M\rangle$ while $\beta<\beta_c(h=0)$. However, for 
$\beta>\beta_c$, there is an spontaneous magnetization 
$\langle M(\beta)\rangle$, and the presence of a magnetic field $h\neq 0$
can only increase $|\langle M(\beta)\rangle|$, therefore values of
$|\langle M\rangle| < |\langle M(\beta)\rangle|$, are not accessible in the
grand canonical formalism. In fact, these non-equilibrium states are separated
from the normal states by the first order transition in $\beta$ mentioned in
section 2.1.

However, we can still consider the values of $\beta$ and the 
magnetization that correspond to non-equilibrium states in the LGM, in the
framework of the model (\ref{model}). 
They will correspond to a definite mean 
value of the energy, for example. These points of the phase diagram are 
points of thermodynamic equilibrium of the model (\ref{model}).
Hence one may ask about the behaviour of the energy
when crossing the corresponding transition point of the LGM and the 
existence of a phase transition at this point in the present model.
We aim to investigate this point in the sequel, as well as
its consequence on the behaviour of the fragment size distributions.

\subsection{Update algorithm}

The Ising model with constant order parameter (magnetization $M$) has been
considered previously from the point of view of its critical dynamic
exponents (see, eg \cite{ZJ}), but it seems that its equilibrium properties
have not been addressed \cite{SH}. The corresponding dynamics would be in
principle that of the Kawasaki spin exchange model \cite{Kaw}, but, as
the authors of \cite{SH} noted, one should take special care with the
detailed balance in a numerical algorithm for this model.

In our Monte Carlo computations, the following update algorithm
has been used: two spins are picked up at random, and they are exchanged 
according to the standard Metropolis prescription \cite{Metro}
only if they have different signs. This algorithm preserves
the magnetization (that is, the number of spins up, which represent the
particles), and it can easily be seen that it satisfies detailed balance.

In fact it is striking to notice how one can arrive to non-Boltzmann
distributions if one does not care very much about the update procedure.
For example, we also considered the following algorithm: take one spin
sequentially on the lattice and another one at random, and exchange them
only if they have different signs. This does not satisfy detailed balance,
as the following argument shows. Consider an initial configuration with every
spin up except for a single spin down. Suppose that the sequential process is 
such that we have to take the spin at the position $i=0$, which is up.
With the algorithm described, there is only a possible final configuration,
the one in which the spin down has been exchanged with the spin up at $i=0$.
If we consider now the reverse process, we have to exchange the spin
down at $i=0$ with a spin up. But now we have many final configurations,
not only the initial one, so it is clear that this process does not
satisfy detailed balance. We can modify this sequential algorithm to 
solve the problem. If the spins taken sequentially and randomly are always
exchanged, independently of their sign, detailed balance is restored. We
verified this numerically, so that the results with this last algorithm were
the same as with the `full randomly' algorithm described above. In fact,
an interesting property of this `random' algorithm in which the two spins are 
taken at random is that here one can choose exchanging the spins in every
case, or only when having different signs. Both situations verify detailed
balance, as we checked numerically, thanks to the randomness in the choice
of the spins.

Similar subtleties happen in algorithms where nearest neighbours
are exchanged, as in Kawasaki dynamics, which were first noted in \cite{SH}.
    
\section{The $(\rho,T)$ phase diagram and \\ cluster formation}

\subsection{Definition of observables}

We simulate the IMFM, defined by eq. (\ref{model}), 
in a three-dimensional cube of
$V=L^3$ sites, and with $N=\rho\cdot V$ spins up and 
periodic boundary conditions.
We work out the phase
diagram in the $(\rho,\beta)$ plane.
 
The thermodynamic observables are the energy
\begin{equation}
E=\frac{1}{3V}\sum_{\langle i,j\rangle}\sigma_i\sigma_j,
\label{energy}
\end{equation}
and its $\beta$ derivative, the specific heat
\begin{equation}
C_v=3V\cdot\left(\langle E^2\rangle - \langle E\rangle^2\right).
\label{cv}
\end{equation}

The droplet observable which we consider
is the second momentum
\begin{equation}
S_2=\sum_s s^2 P(s),
\end{equation}
where $P(s)$ is the number of droplets of size $s$, and the sum extends
over all droplets except for the largest one.

The droplets are defined from the Ising clusters with the probability
law (\ref{bond}). With this definition, the transition properties of the 
thermodynamic transition, given by the divergence of $C_v$, and the
droplet transition, given by the divergence of $S_2$, coincide in the
standard Ising model (constant magnetic field) at $h=0$. 
We want to see whether this is also the case in the IMFM.

\subsection{Phase diagram and cluster formation}

The phase diagram of the IMFM should be the same as
the one of the LGM because the thermodynamic relations 
between the different quantities must be the same in the domain where
the LGM displays an homogeneous phase. The thermodynamic
transition lines must coincide in both phase diagrams. However, the nature
of this transition can be different, as noticed above.

\begin{figure}[tb]
\centerline{\epsfig{figure=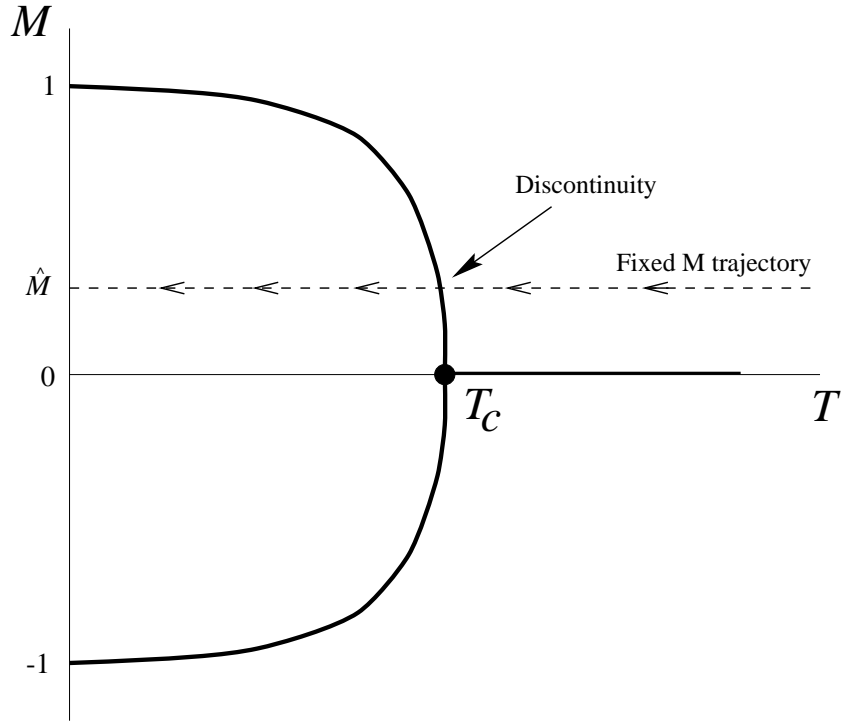,angle=270,width=110mm}}
\caption{Plot of $M(T)$ in the Ising model at $h=0$ magnetic field,
and trajectory of constant magnetization.}
\label{FIG:PHD1}
\end{figure}

We can also understand physically the existence of a phase transition 
in the Ising model with constant magnetization with the help of 
Fig.~\ref{FIG:PHD1}. It shows the curve $M(T)$ for the usual Ising model
at $h=0$ magnetic field. Constraining the system to move in $T$ with
fixed magnetization $\hat M$ (indicated by the dashed line) will be equivalent
to pass through Ising models with different values of $h$ (those which give
the magnetization $M(T)=\hat M$  at each $T$), until one finally reaches the
$M(T)$ curve corresponding to $h=0$. 
Then it is not possible to go to the left in the usual Ising 
model with any magnetic field, but this is possible
with the present model. One expects a phase 
transition at that point, and this will be confirmed below. 
Therefore, the curve $M(T)$ plotted in Fig.~\ref{FIG:PHD1}
will exactly correspond to the transition line of this model 
in the $(\rho,\beta)$ plane.

\begin{figure}[tb]
\centerline{\epsfig{figure=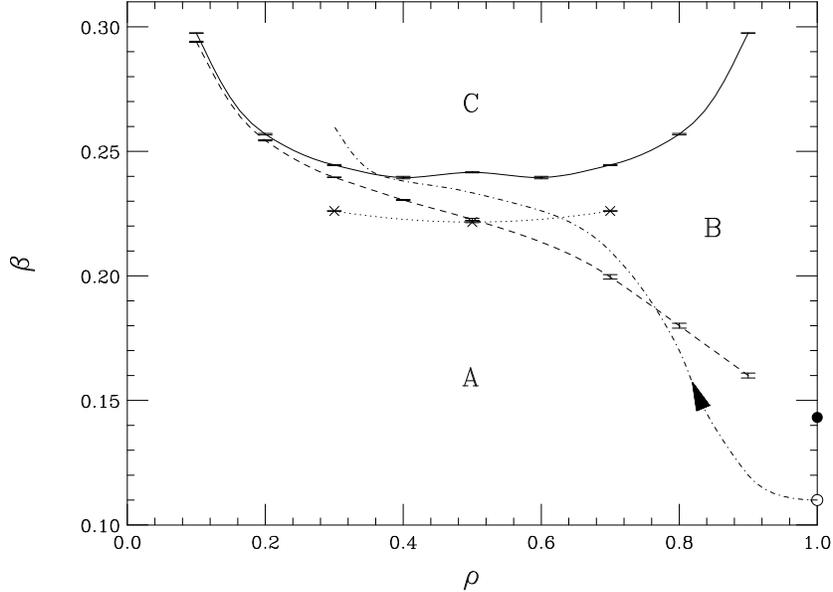,angle=90,width=110mm}}
\caption{Phase diagram of the Ising model with constant magnetization,
obtained numerically from an $L=20$ $3d$ cubic lattice. The solid line is the
thermodynamic transition and the dashed line the droplet transition.
The dotted line indicates the infinite volume limit of the thermodynamic
transition (see text). The black point at $\rho=1$ is the three-dimensional 
bond-percolation point. The line with the arrow (dot-dashed) symbolizes 
a typical trajectory, see text.}
\label{FIG:PHD2}
\end{figure}

In Fig.~\ref{FIG:PHD2} we plot the phase diagram obtained directly in
a numerical simulation of an $L=20$ lattice. We represent the
thermodynamic transition 
(maximum of the specific heat, divergence in the infinite volume limit) 
and the droplet transition (maximum of $S_2$).
For $L=20$ these transitions are different, but a priori this could be
a finite size effect.
We shall see in the next sections
whether these two transitions are the same. The thermodynamic transition
is symmetric with respect to $\rho=0.5$, because the thermodynamic
properties
are the same in terms of spins up or down, that is, the model has a symmetry
$\sigma\leftrightarrow -\sigma$ which corresponds to 
$\rho\leftrightarrow 1-\rho$.
The droplet transition is not
symmetric because we define only droplets for sites corresponding to
$\sigma=+1$. In Fig.~\ref{FIG:PHD2} the thermodynamic transition
for $L=20$ presents an ``anomaly'' at $\rho=0.5$ so that the curve seems
not to be concave there. This is a finite size effect, due to the fact
that the critical exponents vary along the line, as we will see in section
4. Then the thermodynamic limit is reached differently at different points on
the line. Actually, we have included in the Figure the infinite volume curve
obtained from the values of the critical points at $\rho=0.3$ and $\rho=0.5$
calculated in section 4.3.
 
The region A in Fig.~\ref{FIG:PHD2}
is the disordered region, B is a region where an infinite cluster
is formed but the spin system, from a thermodynamic point of view, is
disordered, and C is a region with an infinite cluster which also corresponds
to an ordered thermodynamic system. The thermodynamic transition is the
line which separates the regions B and C, and the droplet transition is the
one which separates the regions A and B.

The physical origin of the infinite cluster is, on the one hand, 
the Ising-type interaction, which, in the nuclear context, simulates 
the strong interaction between the particles. The most probable configuration
for large $\beta$ is that with all spins up together, forming a spherical
cluster. This is only relevant in region C. 
On the other hand, the density also influences the development
of clusters, because we know that, even in absence of interaction, the
system percolates when there are many particles. 
This effect appears in region B, and this is the reason why the 
droplet transition separates from the thermodynamic one for large
values of $\rho$. Besides that, $\beta$ also influences the formation of
physical droplets through the bond probability (\ref{bond}), simulating
a kinetic energy for the particles depending on the temperature: the
lower the temperature, the less the particles can move.

Far inside the ordered phase, the system favours 
the configuration which
minimizes the energy, and then a large cluster
is formed. All occupied sites inside the cluster 
contribute positively to the energy, as well as those 
outside the cluster ($\sigma_i\sigma_j=+1$).
The only contributions with $\sigma_i\sigma_j=-1$ are those
located at the border, so that the configuration will show a minimal 
surface, something very similar to a sphere.
One can estimate the energy in this case.
Let us consider a cubic shape for this
cluster. Then, the total amount of (potential) energy is
\begin{equation}
E_T=3V\left(1-\frac{4\rho^{2/3}}{V^{1/3}}\right),
\end{equation}
so that
\begin{equation}
E=\frac{E_T}{3V}=1-\frac{4\rho^{2/3}}{L},
\end{equation}
and in the thermodynamic limit $(L\to\infty)$, $E$ is normalized to 1.
This is why we chose the factor $3V$ in the normalization of $E$ and
$C_v$ in eqs. (\ref{energy}) and (\ref{cv}). 

\subsection{Cluster distribution and nuclear fragmentation}

Along the droplet transition, the observable $S_2$ diverges, because there
the fragment size distribution $P(s)$ follows a power law
\begin{equation}
P(s)\sim s^{-\tau}.
\label{tau}
\end{equation}


A possible fragmentation scenario can be suggested in the $(\rho,\beta)$
diagram of Fig.~\ref{FIG:PHD2} (dot-dashed line). 
Consider a nuclear collision leading to an excited system
of $N$ particles in a volume $V$, in thermodynamic equilibrium or at least
close to it. This point would correspond to a rather large
density ($\rho\approx 1$)
and small $\beta$ (large temperature), lying below the droplet
transition line. The trajectory would then move towards decreasing 
$\rho$ and increasing $\beta$. It would cross the critical droplet line at some
point $(\rho_0,\beta_0)$ and pursue its way towards lower density and
temperature, up to some freeze-out where the fragment size distribution
gets fixed. From thereon the system would continue to cool down and
expand continuously, crossing the thermodynamic transition line at some
other point. In practice, of course, the system follows a whole set of
trajectories corresponding to different experimental initial
conditions (system, energy, impact parameter,...). Unfortunately it is not
possible to get direct information about these trajectories. There could
exist indications that the crossing of the critical droplet line may occur
for rather high values of the density. In fact, indirect experimental
information (agreement with percolation results) indicates that the exponent
$\tau$ could lie close to the value 2.2 which corresponds to $\rho>0.5$ (see
section 4.4 below). On the other hand, molecular dynamics (MD) calculations
seem to indicate that the system breaks up rather early during the
collision process \cite{Dorso}. 
Whether the critical $\tau$ can be put in direct correspondence with
experimental measurements of fragment distributions which correspond to
averages over different events 
remains however an unsettled question.

\section{Phase transitions and their characteristics}
   
\subsection{Critical exponents}

As we have seen, the phase diagram of the IMFM shows, in principle, 
two phase transitions. We may ask whether
these two transitions coincide for 
small values of $\rho$, 
the type of thermodynamic transition and what is
its order, its universality class, and the value of the exponent $\tau$,
whether it is constant or not along the transition line.

The critical exponents will bring an answer to these questions.
We have two observables which diverge at their corresponding transitions
in the thermodynamic limit, the specific heat
\begin{equation}
C_v\sim (\beta-\beta_c^\mathrm{T})^{-\alpha},
\end{equation}
which defines the exponent $\alpha$, and the second momentum
\begin{equation}
S_2\sim (\beta-\beta_c^\mathrm{D})^{-\bar\gamma},
\end{equation}
which defines the exponent $\bar\gamma$, $\beta_c^\mathrm{T}$ and
$\beta_c^\mathrm{D}$ correspond to the thermodynamic and droplet
transition respectively. Note that in principle, the exponents of the
thermodynamic and droplet transitions will be independent, because we
have two different models: Ising (with fixed magnetization), and a 
bond-percolation correlated model, constructed on the IMFM system.
We know that in the case of the standard Ising model, the definition
(\ref{bond}) of the bond probability is such that the thermodynamic and
the droplet exponents are the same, and the position of the transitions
coincide. But this could be different in the present case. Actually, the
exponent $\gamma$, defined as the one which gives the divergence of the 
magnetic susceptibility, is not defined for the IMFM, because 
we keep the magnetization fixed as an external parameter. This is also
the case of the remaining ``magnetic'' exponents, $\beta$ and $\eta$.
Therefore, the universality class of the IMFM will be completely determined
by $\alpha$ and $\nu$, which should obey the hyperscaling relation
\begin{equation}
\alpha=2-\nu d.
\label{hyp}
\end{equation}
We will denote the exponents corresponding to the droplet transition with
a little bar above the corresponding greek letter.

We can measure critical exponents from standard finite size scaling
\cite{Brezin,Barber}. For finite lattice size $L$, $C_v$ and $S_2$
show peaks (at different points, $\beta_c^\mathrm{T}(L)$ and
$\beta_c^\mathrm{D}(L)$, respectively), which scale with $L$ as
\begin{eqnarray}
C_v(L)&=&A+BL^{\alpha/\nu}, \label{peakcv} \\
S_2(L)&=&\bar A L^{\bar\gamma/\bar\nu}.
\label{peaks2}
\end{eqnarray} 
The exponent $\nu$ can also be obtained from a fit to the law
\begin{equation}
\beta_c(L)-\beta_c(\infty)=AL^{-1/\nu},
\label{nufit}
\end{equation}
where $\beta_c(\infty)$ is the thermodynamic value of the transition
point, and $\beta_c(L)$ is the value of $\beta$ where the thermodynamically
divergent observable has a maximum at finite $L$. 

The exponents of the droplet transition will also obey the
corresponding scaling and hyperscaling relations
\begin{eqnarray}
\frac{\bar\gamma}{\bar\nu}&=&2-\bar\eta,
\label{eta} \\
\bar\alpha+2\bar\beta+\bar\gamma&=&2, \\
\bar\alpha&=&2-\bar\nu d,
\end{eqnarray}
where these exponents are defined by the behaviour of various
moments of the cluster size distributions, see eg \cite{Stauffer}.

Using scaling and hyperscaling with $d=3$, 
the exponent $\tau$ of the distribution of
fragments can be obtained directly from $\bar\eta$:
\begin{equation}
\tau=2+\frac{\bar\beta}{\bar\beta+\bar\gamma}=2+\frac{1+\bar\eta}{5-\bar\eta}=
2.2+\frac{6}{25}\bar\eta+\mathrm{o}(\bar\eta^2),
\label{tau-eta}
\end{equation}
so that if $\bar\eta$ is small (which is usually the case), $\tau$
is $2.2$ modified by a small amount. We remark that for all the models
with small $\bar\eta$ (and for all the standard models in three dimensions,
$\bar\eta\leq 0.1$), $\tau$ can only vary between 2.1 and
2.3. The only nontrivial result is the deviation from 2.2.

We shall be able to check the validity of these relations through the
determination of $\tau$ by means of two independent methods, directly,
by measuring the droplet distribution followed by a fit to eq. (\ref{tau}), and
by computing the critical exponents with the use of eq. (\ref{tau-eta}).

\subsection{The numerical simulation}

In order to compute the critical exponents, and to see if the
droplet and thermodynamic transitions coincide in the infinite volume
limit, we have studied the transitions with large statistics at 
three different values of $\rho$ in the phase diagram, 
$\rho=0.3,0.5,0.7$.
The peak in $C_v$ (eq. (\ref{peakcv})) can be easily located with the spectral
density method \cite{FS}, but unfortunately, this is not possible with $S_2$,
because we construct this observable with bond
probabilities for each configuration.
This means that the value of $S_2$ is not completely determined
by the configuration, and then it is not possible to use the spectral 
density method to extrapolate to different values of $\beta$.
Therefore, we have obtained the 
maximum of $S_2$ with less precision than in the case of $C_v$.

We studied the thermodynamic transition at $\rho=0.3$ and $\rho=0.5$
(because of the symmetry of this transition, $\rho=0.7$ is equivalent
to $\rho=0.3$), in order to fix the exponents $\alpha$ and $\nu$, which inform
us about the universality class of the IMFM model.
In Table~\ref{statistics} we show the lattice sizes and statistics used,
together with the integrated autocorrelation time \cite{Sokal}
for the energy.

For the droplet transition we only studied lattices up to $L=24$, first
because of the difficulty of finding the peak of $S_2$ as mentioned above,
and second because the exponent $\tau$ may be a meaningful physical quantity
in the case of a finite system, which may eventually be directly or
indirectly confronted with the experiment.

In the simulation we proceed as follows: we make 10-15  update
steps (in each step the algorithm described in section 2.4
is applied $V$ times), and then we perform a measurement. At this moment 
the droplets are defined according to (\ref{bond}). Hence we can 
calculate $S_2$.
The errors are computed with the jackknife method \cite{JACK}.


\begin{table}[tb]
\begin{center}
\begin{tabular}{|r|c|c||c|c|}\cline{2-5}
\multicolumn{1}{r|}{} & \multicolumn{2}{c||}{$\rho=0.3$} &
\multicolumn{2}{c|}{$\rho=0.5$}\\ \hline
$L$ & $\tau(E)$ & Sweeps $[\times 10^3\tau(E)]$ 
& $\tau(E)$ & Sweeps $[\times 10^3 \tau(E)]$ \\
\hline \hline
10 & 8.55(3) & 1091 & 12.15(6) & 985 \\
\hline
16 & 12.84(11) & 932 & 22.1(3) & 407 \\
\hline
20 & 15.72(12) & 381 & 27.5(3) & 345 \\
\hline
24 & 20.84(12) & 1061 & 35.2(2) & 680\\
\hline
28 & 27.2(2) & 527 & 43.7(4) & 738 \\
\hline
32 & 33.0(3) & 362 & 53.5(7) & 558 \\
\hline
40 & 63.9(1.1) & 110 & 75.3(8) & 189 \\
\hline
48 & 109(2) & 108 & 110(7) & 43 \\
\hline
\end{tabular}
\end{center}
\caption{Total number of sweeps or update steps 
  (measurements are performed every 10-15 sweeps) and 
  the corresponding integrated
  correlation times $\tau$ for $E$, used in the analysis 
  of the thermodynamic transition at $\rho=0.3$ and $\rho=0.5$.}
\label{statistics}
\end{table} 

\subsection{The thermodynamic transition}

\begin{figure}[tb]
\centerline{\epsfig{figure=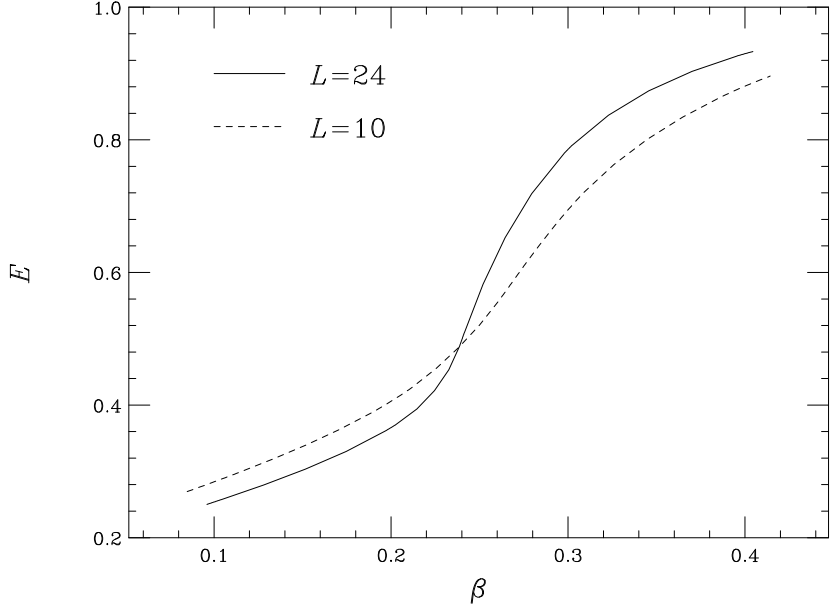,angle=90,width=110mm}}
\caption{Caloric curve ($E$ vs. $\beta$) for two finite lattice sizes,
$L=10$ and $L=24$, at $\rho=0.3$.}
\label{FIG:CCURVE}
\end{figure}

We start by considering the nature of the thermodynamic phase transition
(solid line in Fig.~\ref{FIG:PHD2}),
ie the points in the phase diagram where the specific heat diverges.
We show in Fig.~\ref{FIG:CCURVE} the caloric curve ($E$ vs. $\beta$) 
for two lattice sizes, $L=10,24$. In principle, these kind of curves are not
very precise to distinguish the order of the phase transition. A first
order transition means a discontinuity in this curve, and a second order
one means a vertical slope at the critical line. But this is only true
in the thermodynamic limit, $L=\infty$. 
In a finite lattice there are not singularities at all, and
then it is necessary to carry out a careful study of the thermodynamic
limit by considering finite size scaling techniques, such as expressions
(\ref{peakcv}) and (\ref{peaks2}). The values of the critical exponents
obtained in this way inform us about the order of the phase transition,
and define the universality class in the case of a second order transition.
We will show evidences of
the second order nature of the transition line at every value of $\rho$.
On the one hand, a single peak in the histogram of the energy was observed in
all numerical simulations. This is a strong indication of a second order phase
transition. On the other hand, this was confirmed by an analysis of the
critical exponents $\nu$ and $\alpha$, which show a different behaviour from
that of a first order transition ($\alpha=1$, $\nu=1/d=1/3$). 

Exponents $\nu$ and $\alpha$ determine the universality class of
the transition.
In order to see the values of these exponents, we have
to find the behaviour in the $V\to\infty$ limit. We consider
lattice sizes up to $L=48$ 
and study these exponents by finite size
scaling at $\rho=0.3$ and $\rho=0.5$. 

In Table~\ref{termo}, we report the values of the thermodynamic
transitions and the maxima of $C_v$ for the different
sizes at $\rho=0.3$ and $\rho=0.5$.

\begin{table}[tb]
\begin{center}
\begin{tabular}{|r|c|c||c|c|}\cline{2-5}
\multicolumn{1}{r|}{} & \multicolumn{2}{c||}{$\rho=0.3$} & 
\multicolumn{2}{c|}{$\rho=0.5$}\\ \hline
$L$ & $\beta_c^\mathrm{T}$ & $C_v^\mathrm{max}$ & 
$\beta_c^\mathrm{T}$ & $C_v^\mathrm{max}$ \\
\hline\hline
10 & 0.2698(3) & 3.063(5) & 0.2778(10) & 3.880(7) \\
\hline
16 & 0.2516(3) & 5.109(9) & 0.2495(4) & 6.380(14) \\
\hline
20 & 0.2445(2) & 6.191(8) & 0.2417(2) & 7.739(11) \\
\hline
24 & 0.23988(11) & 7.033(7) & 0.23694(6) & 8.819(9) \\
\hline
28 & 0.23658(11) & 7.758(8) & 0.23364(7) & 9.726(11) \\
\hline
32 & 0.23435(11) & 8.45(2) & 0.23157(8) & 10.519(12) \\
\hline
40 & 0.23152(5) & 9.72(2) & 0.22862(7) & 11.80(3) \\
\hline
48 & 0.229885(14) & 11.11(4) & 0.22680(8) & 12.93(6) \\
\hline
\end{tabular}
\end{center}
\caption{Thermodynamic transition $\beta_c^\mathrm{T}$ and the maxima of
  $C_v$ for different lattice sizes at $\rho=0.3$ and $\rho=0.5$.}
\label{termo}
\end{table} 

\begin{figure}[tb]
\centerline{\epsfig{figure=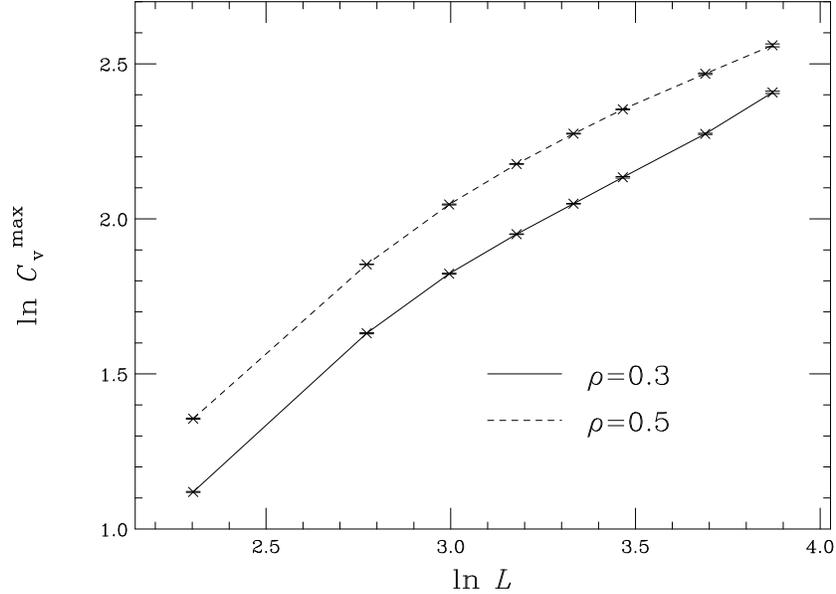,angle=90,width=110mm}}
\caption{Plot of $\ln C_v^\mathrm{max}$ vs. $\ln L$ for 
$L=10,16,20,24,28,32,40,48$ at $\rho=0.3$ and $\rho=0.5$. 
A straight line fit should give $\alpha/\nu$. We observe scaling 
corrections for small $L$ in both cases.}
\label{FIG:SCALING}
\end{figure}

Consider first the case $\rho=0.3$. One needs to make a fit to eq.
(\ref{peakcv}) in order to obtain $\alpha/\nu$. If the constant term is not
very important, a plot $\ln C_v^\mathrm{max}$ vs. $\ln L$ should fit to a
straight line of slope $\alpha/\nu$. This plot is shown in 
Fig.~\ref{FIG:SCALING}, where we see that the slope changes with $L$ for
small $L$. This means that corrections to scaling are important. 
We observe that the slope reaches an asymptotic limit from $L=24$ on.
Actually, the fit for $L=24,28,32,40$ is very good:
\begin{equation}
\alpha/\nu=0.634(4) \quad\quad \chi^2/\mathrm{DF}=0.290/2,
\label{fitanu}
\end{equation}
where DF stands for the number of degrees of freedom in the fit.
The $L=48$ point, however, stays a bit away of this value. In fact, if we
try to include this point, the fit obtained is not good because we
obtain a large $\chi^2/\mathrm{DF}$, and it gives a
value of $\alpha/\nu=0.643(4)$. It is possible that the failure of this fit
is due to a subestimate of the error in this last point, which has a
large autocorrelation time (see Table~\ref{statistics}). To be cautious,
anyway, we will take as the best estimate for $\alpha/\nu$, the value
\begin{equation}
\alpha/\nu=0.639(8).
\label{anufinal}
\end{equation}

An independent fit from this one is the two-parameter fit 
to eq. (\ref{nufit}),
which gives the
best fit for $\nu=0.60(4)$. Here the error is very large. Considering instead
the result (\ref{anufinal}), hyperscaling (\ref{hyp})
gives for $\nu$ and $\alpha$ the values
\begin{equation}
\nu=0.5496(12) \quad\quad \alpha=0.351(4).
\label{nu03}
\end{equation}
One can check this value of $\nu$ by means of a fit to eq. (\ref{nufit}). 
Taking again the lattices $24\leq L<48$, we obtain
\begin{equation}
\beta_c(\infty,\rho=0.3)=0.22606(13) \quad\quad \chi^2/\mathrm{DF}=0.881/2,
\end{equation}
which is a good fit. 
If we add the $L=48$ lattice, this value is slightly
modified ($\beta_c=0.22590(6)$), and the $\chi^2/\mathrm{DF}$ is still
acceptable (3.5/3). Therefore, we will take the values (\ref{nu03}) as the best
estimate for $\alpha$ and $\nu$ at $\rho=0.3$.


Next, consider the case $\rho=0.5$. Here a plot $\ln C_v^\mathrm{max}$ 
vs. $\ln L$ does not give a good fit up to $L=40$, see Fig.~\ref{FIG:SCALING}, 
but if we include the simulation for $L=48$, it seems that an asymptotic
behaviour is reached for the three last lattices $L=32,40,48$. The fit gives 
\begin{equation}
\alpha/\nu=0.513(10) \quad\quad \chi^2/\mathrm{DF}=0.05/1,
\end{equation}
which, using hyperscaling, means
\begin{equation}
\nu=0.569(2) \quad\quad \alpha=0.292(6).
\end{equation}
A fit to eq. (\ref{nufit}) (using the three last lattice sizes) 
with this value of $\nu$ is not bad. It gives
\begin{equation}
\beta_c(\infty,\rho=0.5)=0.2223(3) \quad\quad \chi^2/\mathrm{DF}=0.895/1,
\end{equation}
However, this value of $\beta_c(\infty)$ is not compatible with the one
we had expected, the Ising value $\beta_c(\infty)\sim 0.22165$, though
it is not far from it.
In fact, if we consider the Ising value of exponent $\nu\sim 0.63$, and
use it to obtain $\beta_c(\infty)$ in the fit to eq. (\ref{nufit}),
we obtain quite a good fit with a value of $\beta_c(\infty)$ compatible
with the Ising one:
\begin{equation}
\beta_c(\infty)=0.2216(2) \quad\quad \chi^2/\mathrm{DF}=0.274/1,
\end{equation}
This means that the behaviour observed for $L=32,40,48$ can be still transitory
and we should not discard the three-dimensional Ising values at 
$\rho=0.5$.

The exponents which we determined were obtained with an accuracy which
is high enough to conclude that their variation along the transition
line is soft. This requested a
substantial numerical effort in the framework of the canonical ensemble.
Working in the framework of the microcanonical ensemble may request less
numerical investments. A more detailed study, using different values of
$\rho$, should be done in order to clarify the universality classes along 
the transition line.


\subsection{The droplet transition}

We consider now the droplet transition, ie the points of the
phase diagram where $S_2$ diverges, which means that a droplet percolates
through the system. For $\rho\leq 0.5$ this transition line is very close
to the thermodynamic one, but for $\rho>0.5$, it seems to be very different,
and strictly under the thermodynamic line, which is easy to understand
because it is the high density of particles and not the interaction
which drives the system to percolation in this case.
One expects that for $\rho\leq 0.5$, the two transition lines coincide.
We can check this, as well as the value of exponent $\tau$ 
at different points of the droplet line. 

In Table~\ref{droplet}, we report the $\beta_c^\mathrm{D}(L)$ values 
of the droplet transition
and the maxima of $S_2$ for different lattice sizes at $\rho=0.3$, $\rho=0.5$
and $\rho=0.7$.

\begin{table}[tb]
\begin{center}
\begin{tabular}{|r|c|c||c|c||c|c|}\cline{2-7}
\multicolumn{1}{r|}{} & \multicolumn{2}{c||}{$\rho=0.3$} & 
\multicolumn{2}{c||}{$\rho=0.5$} & \multicolumn{2}{c|}{$\rho=0.7$}\\ \hline
$L$ & $\beta_c^\mathrm{D}$ & $S_2^\mathrm{max}$ & 
$\beta_c^\mathrm{D}$ & $S_2^\mathrm{max}$ & 
$\beta_c^\mathrm{D}$ & $S_2^\mathrm{max}$ \\
\hline\hline
10 & 0.2560(5) & 5.149(2) & 0.225(3) & 8.62(2) & 
0.194(2) & 11.70(5)\\
\hline
16 & 0.2440(2) & 9.402(12) & 0.2230(5) & 20.36(4) &
0.1984(2) & 30.43(5)\\
\hline
20 & 0.23960(10) & 11.91(2) & 0.2227(4) & 30.58(4) &
0.1990(6) & 48.15(5) \\
\hline
24 & 0.2367(3) & 14.16(2) & 0.2227(2) & 42.98(4) & 
0.2005(6) & 70.6(2) \\
\hline
\end{tabular}
\end{center}
\caption{Droplet transition $\beta_c^\mathrm{D}$ and the maxima of
  $S_2$ for different lattice sizes at $\rho=0.3$, $\rho=0.5$ and $\rho=0.7$.}
\label{droplet}
\end{table} 

A fit of the maxima of $S_2$ to eq. (\ref{peaks2}) gives the exponent
$\bar\gamma/\bar\nu$, 
and therefore, $\bar\eta$ from eq. (\ref{eta}), which will allow
us to obtain $\tau$ through eq. (\ref{tau-eta}).

We use only data from lattices of sizes $L=10,16,20,24$, see 
Table~\ref{droplet}, because of the difficulty of locating the droplet
transition (as mentioned before, it is not possible to use the spectral
density method in this case). With these sizes, we are not able to obtain 
a stable value of $\bar\gamma/\bar\nu$ which can be extrapolated 
to the infinite 
volume. We will simply obtain estimates for the exponent $\tau$ which will
depend on the lattice size. These are shown in Table~\ref{gammanu}.


\begin{table}[tb]
\begin{center}
\begin{tabular}{|r|c|c||c|c||c|c|}\cline{2-7}
\multicolumn{1}{r|}{} & \multicolumn{2}{c||}{$L=10,16$} & 
\multicolumn{2}{c||}{$L=16,20$} & \multicolumn{2}{c|}{$L=20,24$}\\ \hline
$\rho$ & $\bar\gamma/\bar\nu$ & $\tau$ &
$\bar\gamma/\bar\nu$ & $\tau$ &
$\bar\gamma/\bar\nu$ & $\tau$ \\ 
\hline\hline
0.3 & 1.281(3) & 2.4015(12) & 1.060(9) & 2.478(3) &
0.949(12) & 2.519(5) \\
\hline
0.5 & 1.829(6) & 2.242(3) & 1.823(11) & 2.244(3) &
1.867(9) & 2.233(2) \\
\hline
0.7 & 2.034(10) & 2.192(2) & 2.056(9) & 2.188(2) &
2.10(2) & 2.176(5) \\
\hline
\end{tabular}
\end{center}
\caption{Values of $\bar\gamma/\bar\nu$ and $\tau$, taking the data from 
Table~\ref{droplet}. We observe the evolution with $\rho$ for
every lattice size.}
\label{gammanu}
\end{table} 

\begin{figure}[tb]
\centerline{\epsfig{figure=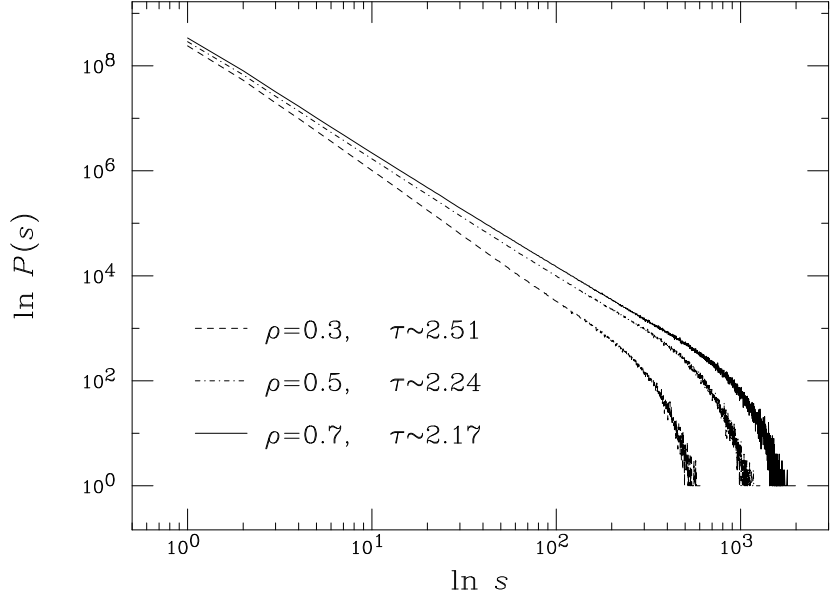,angle=90,width=110mm}}
\caption{Number of droplets $P(s)$ with $s$ particles in logarithmic
scale. The slopes of the fitted straight lines give exponent $\tau$. This is
done for $L=20$ at $\rho=0.3,0.5,0.7$.}
\label{FIG:TAU}
\end{figure}

We can check now the values of $\tau$ with the second method: directly
from the histogram of the
number of droplets of a certain size. In Fig.~\ref{FIG:TAU} we show some of
these histograms, for $L=20$ at $\rho=0.3$, $\rho=0.5$ and $\rho=0.7$,
in logarithmic scale. The actual
value of the slopes are not very precise, because they depend on the region
where the fit is performed. 
Nevertheless, there are central zones in every histogram which
give slopes stable under small variations of their length, and where
we retrieve the values of $\tau$ obtained in Table~\ref{gammanu}.
This means that scaling and hyperscaling are satisfied, and we see that
the results show a dependence of exponent $\tau$ on $\rho$ (besides the
finite size dependence, of course), which suggests a continuous change of
the values of this exponent along the droplet transition line.

Finally, one may raise the question of whether the transitions 
merge together at
the thermodynamic limit. We show the 
differences between $\beta_c^\mathrm{T}$ and $\beta_c^\mathrm{D}$ at
$\rho=0.3,0.5,0.7$ for the lattice sizes $L=16,20,24$  
in Table~\ref{deltabeta}. 

\begin{table}[tb]
\begin{center}
\begin{tabular}{|r|c|c|c|}\hline
$L$ & $\Delta\beta(0.3)$ & $\Delta\beta(0.5)$ & $\Delta\beta(0.7)$\\ 
\hline\hline
16 & 0.0076(4) & 0.0265(6) & 0.0532(4) \\ \hline
20 & 0.0049(2) & 0.0190(4) & 0.0455(6) \\ \hline
24 & 0.0032(3) & 0.0142(2) & 0.0394(6) \\
\hline
\end{tabular}
\end{center}
\caption{Differences between the thermodynamic and the droplet transition
points at $\rho=0.3,0.5,0.7$.}
\label{deltabeta}
\end{table} 

If we try a fit of the data of Table~\ref{deltabeta} to 
$\Delta\beta=AL^{-x}$, we obtain:
\begin{eqnarray}
\rho=0.3& \quad x=2.1(2) &\quad x^{-1}=0.48(5), \quad\chi^2/DF=0.28/1 
\label{fit03} \\
\rho=0.5&  \quad x=1.54(6)& \quad x^{-1}=0.65(3), \quad\chi^2/DF=0.19/1
\label{fit05} \\
\rho=0.7&  \quad x=0.73(4) &\quad x^{-1}=1.37(7), \quad\chi^2/DF=0.31/1 
\label{fit07} 
\end{eqnarray}

We see that the fits (\ref{fit03}) and (\ref{fit05}) for $\rho=0.3$ and
$\rho=0.5$ respectively, are quite good. One 
could argue that if exponents $\nu$ and $\bar\nu$ were the same, this value
should be equal to $x^{-1}$ in the fit above. In fact we see that in 
(\ref{fit03}), the value of $x^{-1}$ is near to (although not completely
compatible with) the value of $\nu$ that we obtained in eq. (\ref{nu03}).
In the case $\rho=0.5$, we obtain a value of $x^{-1}$ which is compatible
with the $3d$-Ising $\nu$ exponent (0.63). This agrees with our conclusion
at the end of the last section that a compatibility with the Ising exponents
should not be discarded at $\rho=0.5$.

Finally, we see from (\ref{fit07}) that a fit to $\Delta\beta=0$ in the
$L\to\infty$ limit is also possible with these data. However, the value of
$x$ is now less than $1$ and has no relation with exponent $\nu$. We know
that the droplet transition line must end at the bond-percolation point
(see Fig.~\ref{FIG:PHD2}), so that we expect that this fit is a spurious
effect, or, equivalently, that exponent $x$ in this kind of fit will 
approach zero for larger lattice sizes. We then conclude that for
$\rho>0.5$ the thermodynamic and droplet transitions are different.

\section{Conclusions}

In the present work we presented and analyzed a simple though 
realistic $3d$ Ising-type model with
fixed magnetization (density) which we considered as a generic description
of nuclear matter fragmentation
in thermodynamic equilibrium. We worked out the phase diagram
of the system in terms of density and temperature. We analyzed
the thermodynamic properties of the system as well as the properties of the
size distribution of bound fragments. For compact systems (relative density
larger than 0.5) we found two types of transitions, one corresponding to a
percolation transition for the system which breaks into pieces, and a
thermodynamic transition at lower temperature. For dilute systems (relative
density lower than 0.3), which should correspond to situations where the 
expanding system reaches the freeze-out point, the two transitions merge and 
the exponents related to the thermodynamics and the fragment distributions 
change with the density.

There are strong indications that the scaling and hyperscaling relations 
between the exponents are satisfied. In all cases which were worked out
numerical results indicate
that the transitions can be interpreted as being second order.
For this reason, one cannot expect to 
observe the experimentally controversed
latent heat plateau in the caloric curve, at least not in a clear-cut
way in the small systems which we considered here. One may perhaps argue
here that the energy fluctuations which are present in the canonical
treatment performed here smear out the energy and hence preclude a
clear-cut observation of the plateau. 
We insist however on the fact that the critical exponents
definitely tell us that the transitions are second order 
as already mentioned above even if at the present stage of our investigations
we cannot say with certitude to which universality class the transitions
belong (it seems clear, however, that they experiment a soft variation
along the transition line).
Actually, the fact that the transition is first order for non-fixed
magnetization, and second order when the magnetization is fixed could have
been expected somehow. For a fixed magnetization, the configuration space
is smaller, there are less fluctuations, and it is natural that the
transition gets weaker.

It remains nevertheless true that the fragmenting nuclear system is isolated
(if one considers the \textsl{whole} system which enters the reaction
process leading to its fragmentation) which justifies a microcanonical
treatment as advertised by D. Gross and collaborators \cite{Gross}.
It may be worthwhile to consider this point, although it is not clear to
us whether it would clarify the situation since in the microcanonical
ensemble it is the temperature which fluctuates strongly at the transition 
points for fixed energy, when the reverse is true in the canonical ensemble.
It is our aim to work out the present model
in this framework in the next future.

Last but not least, the present work shows clearly that models which describe
finite size systems and which differ by seemingly harmless characteristics
(strictly fixed number of particles in IMFM here, fixed number of particles
in the average in the LGM) can lead to quite different results as far as
the order of the transition is concerned. Experimental data should be
enough precise and complete in order to be able to distinguish between 
details like the order of the transition. This is certainly a great
challenge and an incentive to improve our information about the properties
of fragmenting nuclei.

\section*{\protect\label{S_ACKNOWLEDGES}Acknowledgements}

We wish to thank P. Wagner, L.A. Fern\'andez, J.J. Ruiz-Lorenzo,
D. \'I\~niguez, J.L. Alonso, A. Fern\'andez-Pacheco, 
M. Flor\'{\i}a and X. Campi for discussions. We also acknowledge
D. Gross for his constructive remarks.
J.M.C. is a Spanish MEC fellow. He also thanks the CAI European
program and DGA (CONSI+D) for financial support.
The numerical work was done using the RTNN parallel machine (composed
by 32 Pentium Pro processors) located at Zaragoza University.

\end{document}